\documentclass[a4paper,10pt]{article}

\usepackage[latin1]{inputenc}
\usepackage[brazil]{babel}
\usepackage[top=2.5cm, left=2.5cm, right=2cm, bottom=2.5cm]{geometry}
\usepackage{graphicx}
\usepackage[centertags]{amsmath}
\usepackage{amsfonts}
\usepackage{amssymb}
\usepackage{amsthm}
\usepackage{newlfont}
\usepackage{graphicx}

\newlength{\defbaselineskip}
\begin{document}

\begin{center}
{\LARGE Teoria Algébrica de Processos da Medida em Sistemas Quânticos}

{\small (Algebraic Theory of Measurement Processes in Quantum Systems)}

\bigskip

C. A. M. de Melo$^{\dagger ,\ddagger ,}$\footnote{%
cassius@unifal-mg.edu.br}, B. M. Pimentel$^{\dagger ,}$\footnote{%
pimentel@ift.unesp.br}, J. A. Ramirez$^{\dagger ,}$\footnote{%
alrabef@ift.unesp.br}

\bigskip

$^{\dagger }$\textit{Instituto de Física Teórica - Universidade Estadual
Paulista, }

\textit{R. Dr. Bento Teobaldo Ferraz, 271 Bloco II - Barra Funda,}

\textit{C.P. 70532-2, 01156-970,\ São Paulo, SP, Brasil.}

$^{\ddagger }$\textit{Instituto de Ciência e Tecnologia - Universidade
Federal de Alfenas, Campus Poços de Caldas, }

\textit{BR 267 - Rodovia José Aurélio Vilela, nº 11.999, Km 533,}

\textit{Cidade Universitária, 37715-400, Poços de Caldas, MG, Brasil.}

\bigskip

\textbf{Resumo}
\end{center}

Neste artigo trataremos, de uma maneira pedagógica, a forma em que pode ser
construída uma estrutura algébrica para os processos de medida em Mecânica Qu%
ântica partindo do conceito de \emph{Símbolo de Medida}, concebido por
Julian S. Schwinger, e que constitui peça fundamental para seu formalismo
variacional e suas diferentes aplicações.

\textbf{Palavras-chave}: Mecânica Quântica, Teoria Quântica da
Medida.\bigskip

\begin{center}
\textbf{Abstract}
\end{center}

Here we deal in a pedagogical way with an approach to construct an algebraic
structure for the Quantum Mechanical measurement processes from the concept
of \emph{Measurement Symbol}. Such concept was conceived by Julian S.
Schwinger and constitutes a fundamental piece in his variational formalism
and its several applications.


\section{Introdução}

A formulação desenvolvida por Julian Seymour Schwinger para a derivação das
amplitudes de probabilidade\footnote{%
Schwinger denominava as amplitudes de probabilidade (ou funções de onda) de 
\emph{funções de transformação}$.$ Os motivos disto serão esclarecidos mais
adiante.} em Mecânica Quântica (M.Q.) provém de um extenso estudo que deu
seu começo com a análise alternativa dos processos de medida associados à
cinemática da Mecânica Quântica, até a formulação do Princípio Variacional
que caracteriza a dinâmica desses processos. Schwinger foi fortemente
influenciado pelos trabalhos de I. I. Rabi entre 1931 e 1939 \cite{rabiint},
que tratavam de experimentos sobre a interação de feixes de núcleos atô%
micos, ou moléculas, com campos magnéticos \cite{climbint}.

No ano de 1955, na conferência de \textit{Les Houches} \cite{Houchesint},
Schwinger expôs sua idéia sobre a construção de uma álgebra partindo dos
resultados nos processos de medida realizados sobre um sistema quântico para
a obtenção de informação. Neste ponto, Schwinger começou pela revisão da
teoria cinemática desses processos em M.Q. \cite{schwingercu}, \cite%
{schwingercu0}, onde a motivação dada por Rabi intervém. Dado que as partí%
culas com spin respondem ante campos magnéticos externos, elas podem ser
separadas por aquela característica em um experimento tipo Stern-Gerlach
(S.G.).

Nos seus experimentos mentais, Schwinger tinha vários arranjos experimentais
do tipo S.G., cada um deles, possuindo a propriedade de fazer a escolha de
uma ca\-rac\-te\-rís\-ti\-ca genérica do sistema. Assim, este era dividido
em sistemas menores, cada um destes com uma característica em princípio bem
definida. \ Desta forma, precisando somente dos resultados das medidas para
a caracterização dos sistemas quânticos, espera-se que a M.Q. possa ser
derivada diretamente de fatos experimentais.

Uma das primeiras exposições pedagógicas sobre a formulação de Schwinger
apareceu em \cite{NotasInternas}. Apresentaremos aqui uma versão mais curta,
ainda que completa, da descrição cinemática da M.Q. na abordagem de
Schwinger.

Na primeira seção deste artigo, falaremos sobre a medida em sistemas clá%
ssicos e veremos como ela adquire importância quando as grandezas no sistema
se fazem suficientemente pequenas. Depois, iremos definir o conceito de 
\emph{Símbolo de Medida }e\emph{\ }construiremos\emph{\ }o arcabouço que
origina os experimentos mentais, permitindo compreender a estrutura matemá%
tica que rege os processos quânticos no ato de medida; assim, partindo da
abstração de tais processos e de como se dá sua incidência sobre o sistema,
se mostrará a importância da história na construção do mesmo. Desta forma, a
análise das medidas consecutivas realizadas sobre um sistema ajudará a
compreender como ele poderia ser preparado e, igualmente, como este poderia
ser caracterizado,\ levando, ao final, a uma importante relação entre os
processos de medida e os fatos de escolha, que são refletidos na interpretaçã%
o estatística das funções de transformação. \ Posteriormente, veremos como
as caracterizações de um sistema, por meio de dois observáveis não
necessariamente compatíveis, podem se tornar equivalentes com o uso de funçõ%
es de transformação e com um tipo especial de transformação chamada de 
\textit{unitária} que, fundamentalmente, preserva a informação que se
possui\ do sistema, como a probabilidade ou a norma de um vetor que
represente um estado em um espaço de Hilbert. Assim, este tipo de transformaç%
ão permite o estudo da cinemática\ quântica nos processos envolvidos.

Por último, construiremos as estruturas convencionais nos processos cinemá%
ticos associados à Mecânica Quântica, e relacionaremos estes com estruturas
num espaço vetorial.

\section{A Medida em um Sistema Físico}

Do ponto de vista clássico, uma medida tem como idealização que o efeito do
a\-pa\-re\-lho de medição sobre o sistema físico de interesse é desprezível
ou que não o afeta de uma maneira apreciável e, se isto ocorrer, tal efeito
pode ser compensado em uma forma estatística ou em uma forma mecânica \cite%
{schwingercu}, \cite{schwingercu0}, \cite{schwingercu1}. Mas a interação
entre o sistema e o aparelho de medida toma importância no momento que
considerarmos sistemas microscópicos; nestes casos, não há interação que
possa ser considerada pequena, o sistema pode ser alterado facilmente, e o
efeito da medida não pode ser compensado por nenhum meio, dado que a
sensibilidade do sistema ante qualquer novo procedimento pode afetá-lo e
fazer com que os resultados obtidos não representem mais os estado anterior à
medida, ou sejam imprevisíveis.

Segundo esta ordem de idéias, podemos observar que a sensibilidade (a
resposta) de um sistema diante de determinados estímulos como a ação de
medida, é baseada no fato de que, para obter informação do sistema, temos
que fazer uma troca de energia com ele. Assim, independentemente da sua
escala, precisamos introduzir energia no sistema e depois observar as formas
em que ele reage; como, por exemplo, emitindo algum tipo de radiação.

Desta forma, a ação de obtenção de informação, a ação de medida, pode ser
simbolizada por meio da quantidade $R$ \cite{john}, que simboliza sua
resposta ante estímulos externos, e que é dada pela razão entre a quantidade
de energia $\Delta E_{m}$\ necessária para induzir uma mudança no sistema e $%
E_{s}$\ a energia total que possui o sistema, de modo que%
\begin{equation}
R=\frac{\Delta E_{m}}{E_{s}}.  \label{r}
\end{equation}%
Dada a forma da construção de $R$, classicamente o menor valor que pode
tomar para $E_{s}$\ fixo é $\min (\Delta E_{m})=0$ e qualquer flutuação
deste valor evidencia uma variação da energia no sistema. Disto podemos
dizer que o estado do sistema pode mudar depois de uma medição, já que sua
energia vai mudar de uma maneira apreciável.\ Entretanto, se levarmos esta id%
éia para sistemas microscópicos, mantendo em conta a quantização da energia,
esta proporção (\ref{r}) deve estar relacionada com a constante de Planck $h$
e, portanto, se obtemos algum tipo de informação do sistema, não pode
existir um valor de $\Delta E_{m}<\alpha h$ \footnote{$\alpha $\ é uma
constante que modifica a expressão para unidades de energia.}, de tal forma
que, no nível microscópico, \textbf{não pode haver nenhuma medida que não
afete o sistema}. Como temos visto, o raciocínio trata sobre a produção de
uma mudança na energia do sistema originada por uma medição e, portanto, o
objetivo da medida em sistemas quânticos será caracterizar o que ocorre no
sistema quando estes procedimentos são efetuados. \ Assim, quantificar de
certa forma a incidência de uma medida sobre um sistema microscópico, ao
estudar os possíveis resultados e as relações que eles têm, será o objetivo
das próximas seções.

\section{Simbologia da Medida}

Quando medimos alguma característica específica de um sistema, por exemplo a
quantidade $A$, podemos em princípio obter uma grande quantidade de
resultados, como $\left\{ a_{1},a_{2},a_{3},a_{4},...\right\} $,
classificados pela sua grandeza. Chamaremos genericamente de \textbf{observá%
vel }à quantidade $A$ com que vai se ca\-rac\-te\-ri\-zar o sistema e ao
conjunto de números $\mathbf{E[}A\mathbf{]=\{}a_{j}\mathbf{\}}$\textbf{,}
com $j$ como índice do conjunto \footnote{%
Nota-se que o conjunto dos índices $j$ ou $k$ com $j,k\subseteq N$, é $%
j=\{0,1,2,...,n\}$; se o espectro for contínuo esta notação deve ser
entendida como fazendo referência a um intervalo nos números reais.},\ o
espectro de $A$. \ Um sistema pode ser descrito por um número infinito de
observáveis, $A$, $B$, $C$, $D$, $E$,\textbf{...}, que terão espectros $%
\mathbf{E[}A\mathbf{]}$\textbf{, }$\mathbf{E[}B\mathbf{]}$\textbf{, }$%
\mathbf{E[}C\mathbf{]}$\textbf{, }$\mathbf{E[}D\mathbf{]}$\textbf{,...,} mas
cada uma destas descrições representará de maneira única o sistema.

\begin{figure}[ht]
                           \begin{center}
                           \includegraphics[width=5 cm, height=2 cm]{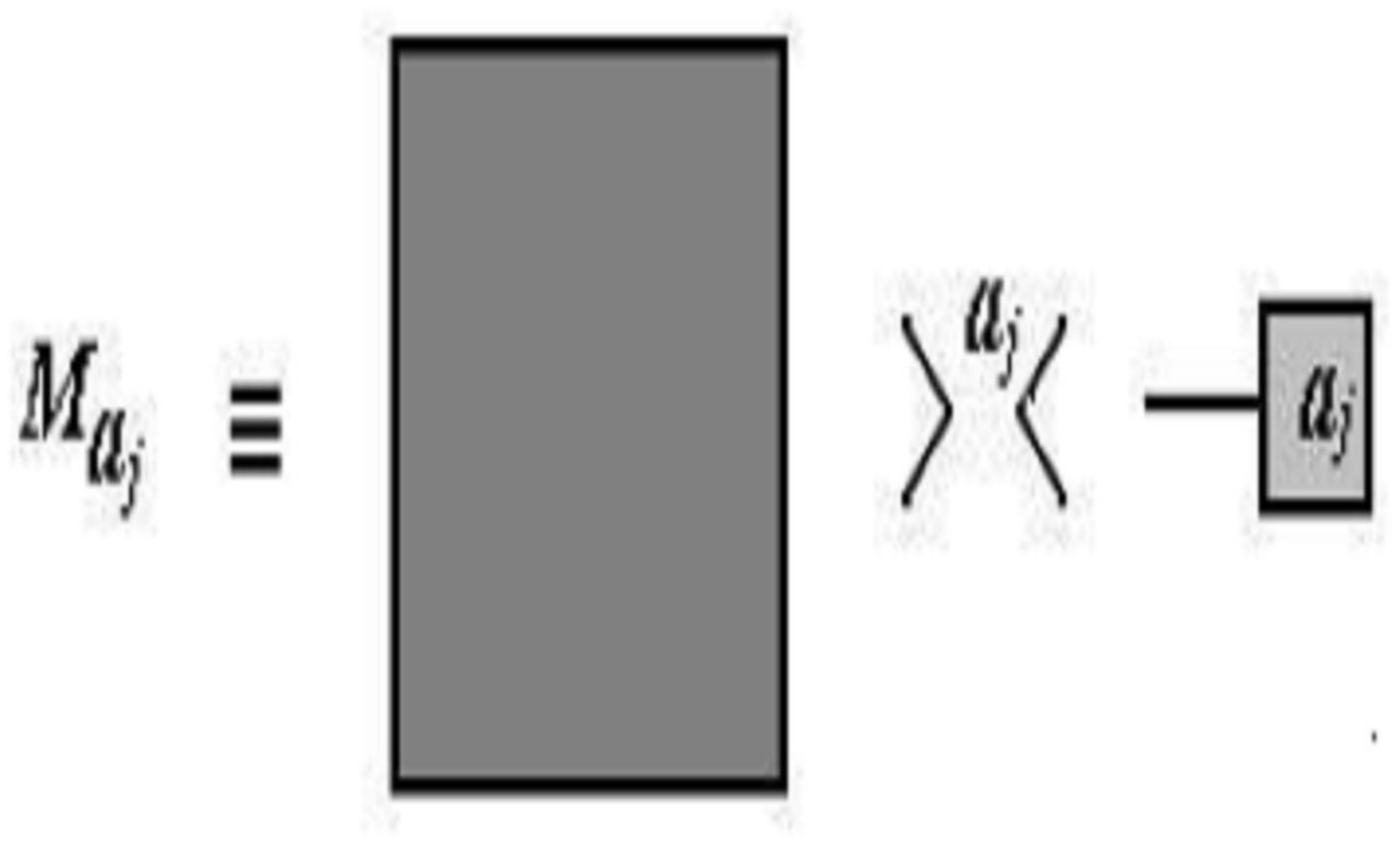}
                           \caption{O feixe ingressante de partículas, sistema, representado pelo quadro cinza-escuro é filtrado extraindo um feixe emergente de partículas no qual os constituentes têm o valor $a_{j}$ de $A$.}
                           \end{center}
                           \label{esquema}
 \end{figure}

Para poder exemplificar um pouco mais as idéias anteriores, suponhamos um
feixe de partículas que pode se considerar como um grande conjunto de
sistemas, em que cada um deles pode ter um valor bem definido $a_{j}$ de $A,$
e onde cada sistema pode se separar do conjunto por um processo de filtragem
tipo Stern-Gerlach. Cada sub-sistema resultante com valor $a_{j}$, se diz
estar no estado $a_{j}$. O processo de medida pode ser ilustrado na Figura %
\ref{esquema}, sobre este tipo de processos construiremos a álgebra de
medida.

As relações entre tais símbolos
fixam uma série de operações que serão a\-na\-li\-sa\-das a seguir.

\section{Relações Entre Medidas Sucessivas}

As relações entre os símbolos de medida expressam a essência dos processos
de seleção. Assim, a cada medida realizada sobre o sistema é assinalado um sí%
mbolo $M_{a_{j}}$\ que representa uma seleção dos estados com o valor $a_{j}$%
\ e rejeita aqueles que se encontram em um estado diferente. Como foi visto
na seção anterior, os símbolos são rotulados pela quantidade que está sendo
medida. Assim, o número de símbolos de medida que temos corresponde ao valor
máximo de estados do conjunto $\left\{ j\right\} $, $\max \left[ j\right] $.
\ As relações seguintes são deduzidas das ca\-rac\-te\-rís\-ti\-cas dadas
para os símbolos de medida e, portanto, expressam somente a forma em que
estes operam em conjunto. \ Dependendo, assim, da forma como são realizadas
as medidas, temos que:

\begin{enumerate}
\item A \textbf{soma} de dois símbolos de medida, representada da seguinte
forma:

\begin{figure}[ht]
                           \begin{center}
                           \includegraphics[width=7 cm, height=2 cm]{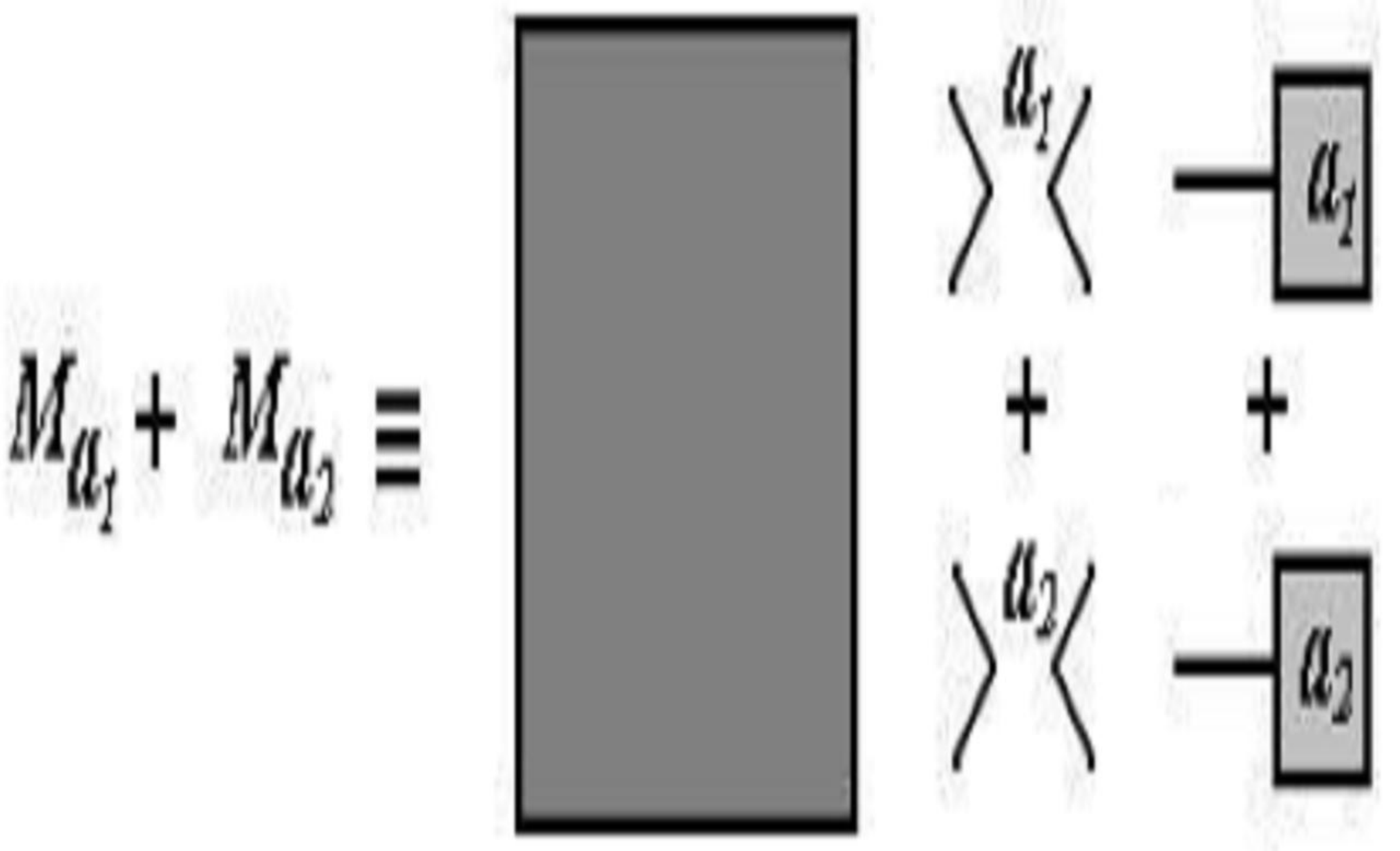}
                           \caption{Esta ação corresponde à medição
que aceita estados que têm o valor $a_{1}$\ ou $a_{2}$\ sem fazer distinção
alguma.}
                           \end{center}
                           \label{fig2}
 \end{figure}
 
Este processo é realizado si\-mul\-ta\-nea\-men\-te, sendo que esta medida é menos seletiva admitindo
sistemas no estado $a_{1}$ ou no estado $a_{2}$ e produz sub-ensembles
associados aos subconjuntos de $\mathbf{E[}A\mathbf{]}$\ para este caso em
especial. Assim, de uma maneira mais geral, também é válida para mais sí%
mbolos de medida onde temos, além disso, associatividade.

\begin{figure}[ht]
                           \begin{center}
                           \includegraphics[width=7 cm, height=3 cm]{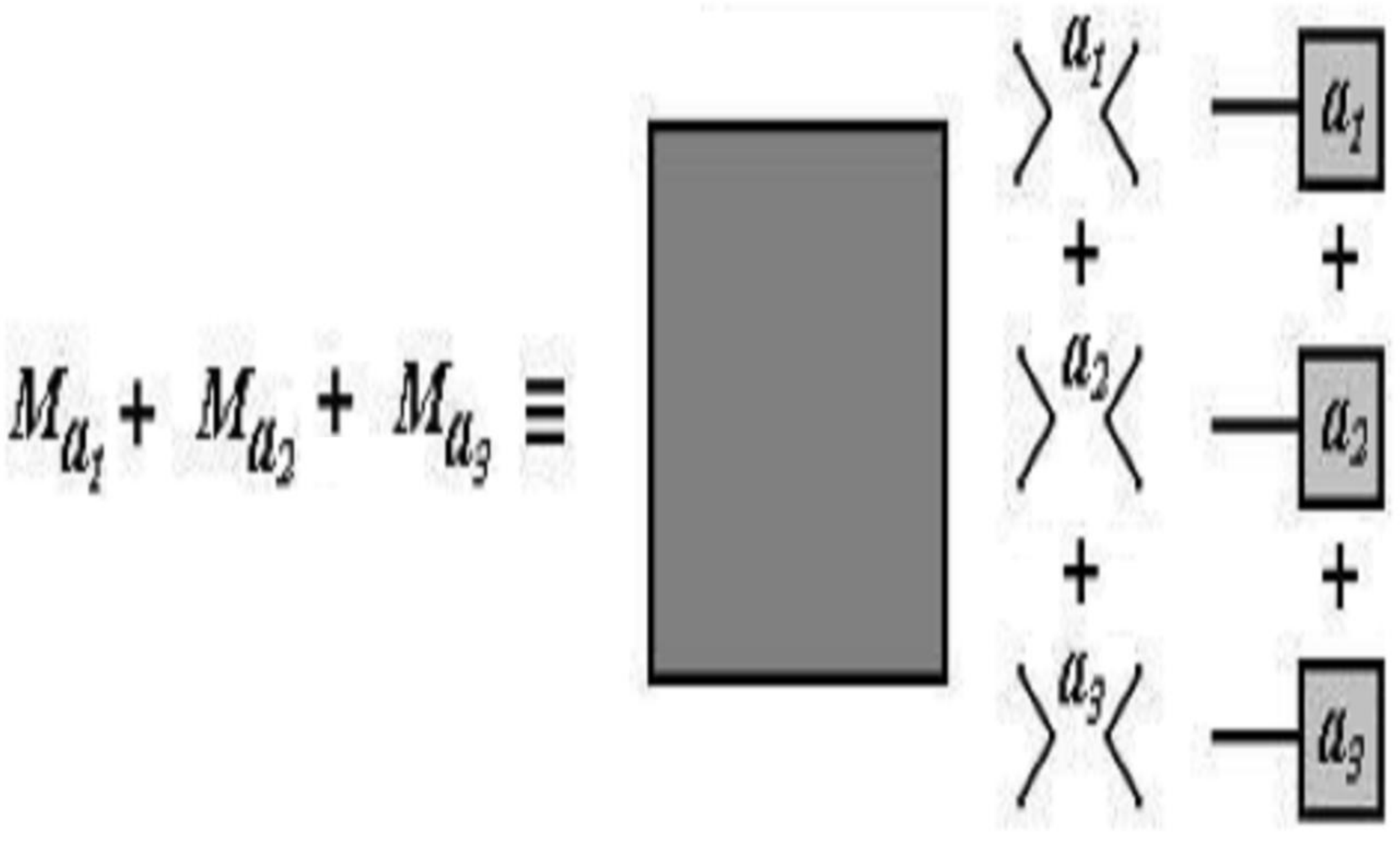}
                           \caption{Neste esquema, se mostra como atua a medida
de três caracteristicas simultâneamente.}
                           \end{center}
                           \label{fig3}
 \end{figure}

Assim obtemos 
\begin{equation}
\left( M_{a_{1}}+M_{a_{2}}\right) +M_{a_{3}}=M_{a_{1}}+\left(
M_{a_{2}}+M_{a_{3}}\right) .
\end{equation}%
Com estas propriedades, podemos inferir duas implicações fundamentais: a
pri\-mei\-ra é assinalar um símbolo para o processo de medição que deixa
passar todos os estados sem distinção, o símbolo para este processo será $%
\mathbf{1}$, que designaremos como Medida Completa, e que pode ser
representado da seguinte forma:

\begin{figure}[ht]
                           \begin{center}
                           \includegraphics[width=9 cm, height=6 cm]{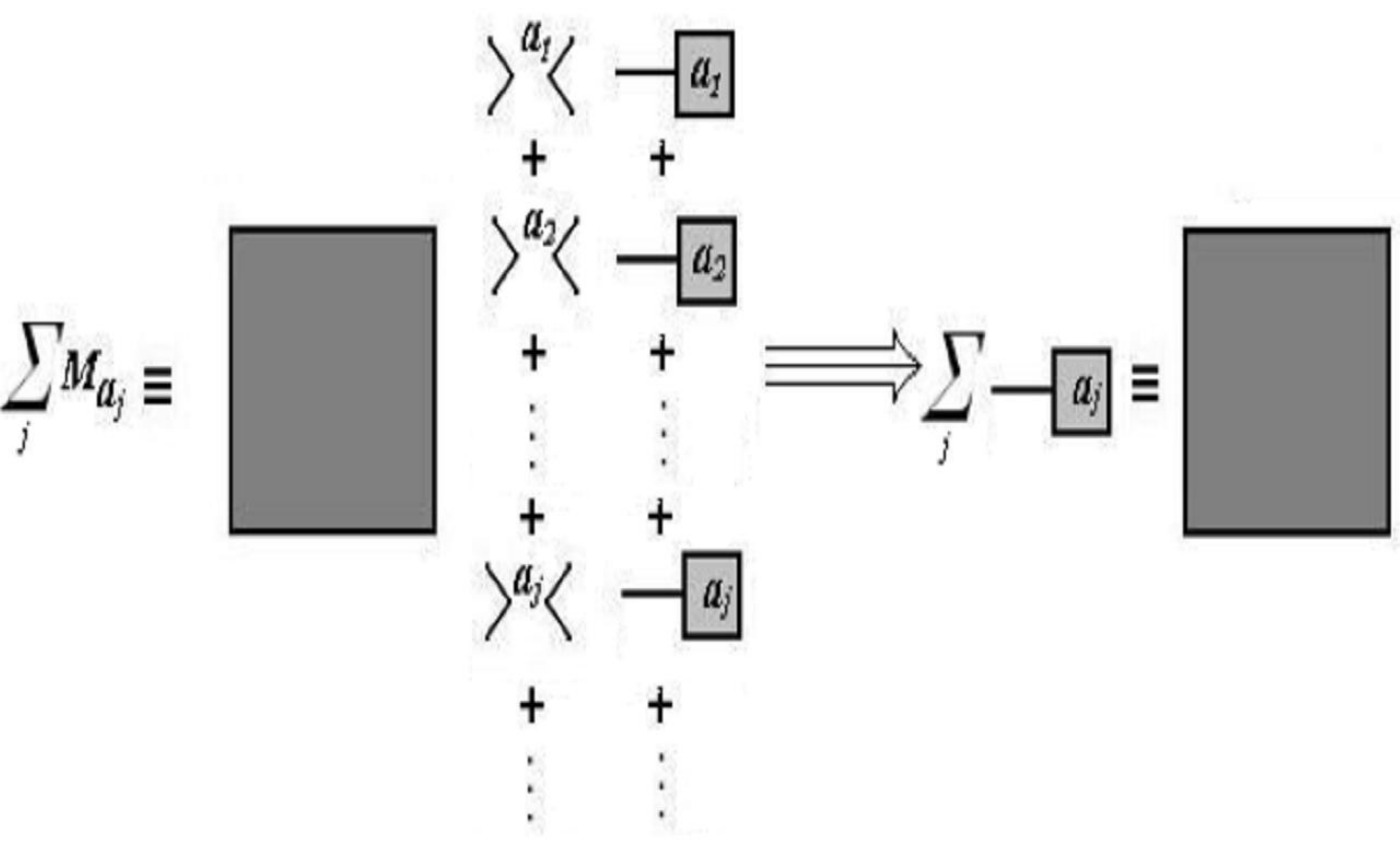}
                           \caption{A
representação da unidade como a soma de todos os simbolos associados a um
observavel, mostra que fazer com que o sistema pase por todos os filtros o
deixa passar inteiro.}
                           \end{center}
                           \label{fig4}
 \end{figure}

Assim, temos que%
\begin{equation}
\sum_{k=0}^{N}M_{a_{k}}=\mathbf{1},  \label{sum}
\end{equation}%
e a segunda é dada por completude, podendo ser definido o símbolo $\mathbf{0}
$\ que significa que há um processo que rejeita qualquer estado.

\item O \textbf{produto} de dois símbolos de medida $M_{a_{1}}M_{a_{2}}$,\ e
cuja leitura é rea\-li\-za\-da da direita para esquerda, e que pode ser
representado da seguinte forma: 

\begin{figure}[ht]
                           \begin{center}
                           \includegraphics[width=9 cm, height=4 cm]{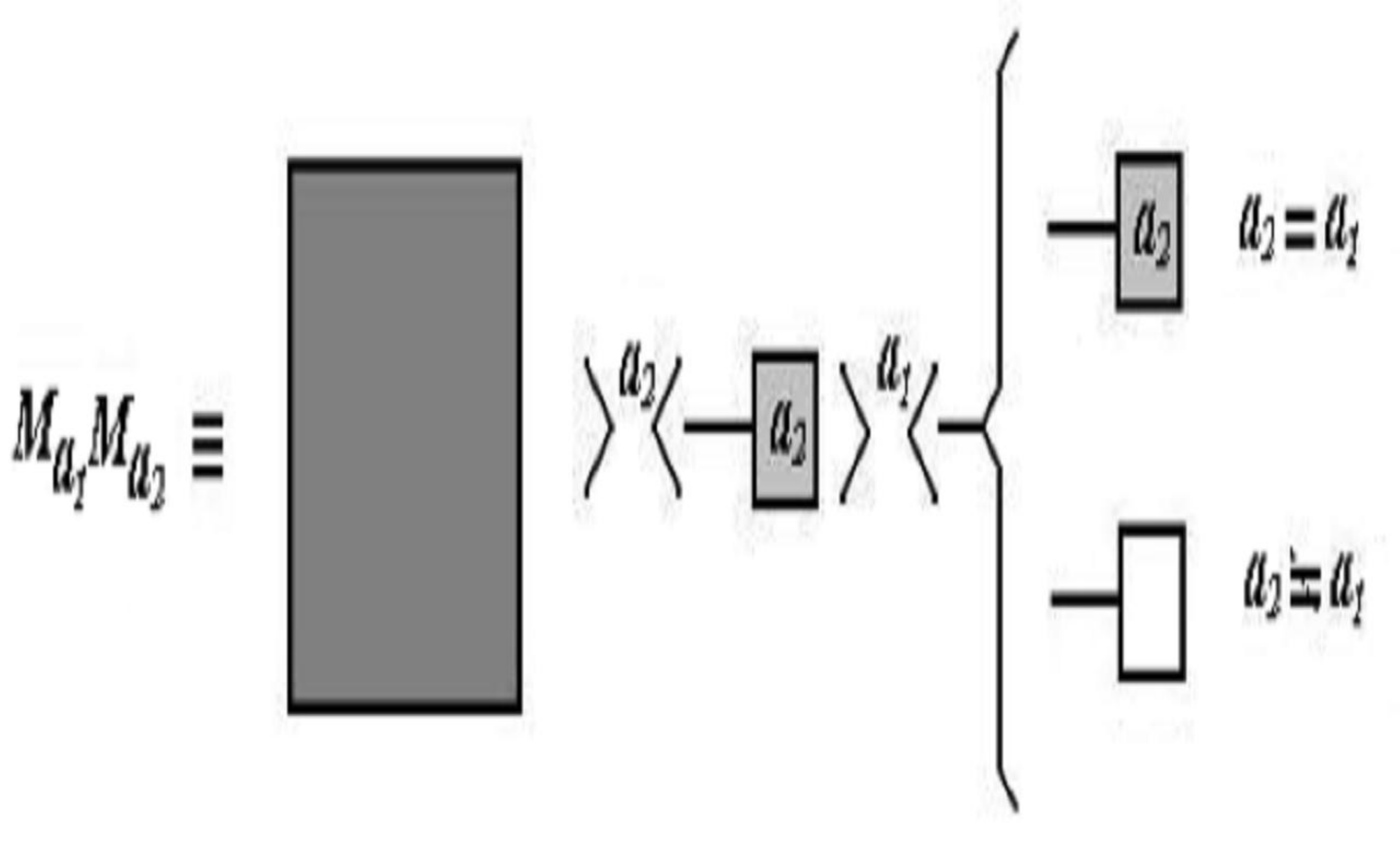}
                           \caption{Este processo simboliza a filtragem de\ todos os sistemas que têm o
valor de $a_{1}$\ dos sistemas emergentes do processo $M_{a_{2}}$,\ se $%
a_{2}\neq a_{1}$ a segunda filtragem vai rejeitar os sistemas provenientes
do primeiro processo, assim o resultado será nulo e poderá ser caracterizado
pelo símbolo $0$. \ Entretanto, se $a_{2}=a_{1}$\ o resultado do processo nã%
o é mais nulo.}
                           \end{center}
                           \label{fig5}
 \end{figure}

Portanto, podemos por meio da seguinte quantidade 
\begin{equation}
\delta (a_{2},a_{1})=\left\{ 
\begin{array}{c}
1\mbox{ \ \ \ \ \ \ \ \ }a_{2}=a_{1} \\ 
0\mbox{ \ \ \ \ \ \ \ \ }a_{2}\neq a_{1}%
\end{array}%
\right. ,  \label{delt}
\end{equation}%
representar os resultados anteriores como%
\begin{equation}
M_{a_{1}}M_{a_{2}}=\delta (a_{2},a_{1})M_{a_{2}},
\end{equation}%
podemos ver que temos%
\begin{equation}
M_{a_{2}}M_{a_{2}}=M_{a_{2}},
\end{equation}%
o que mostra que estes símbolos são idempotentes.
\end{enumerate}

As relações anteriores permitem identificar os $M_{a}$ com elementos de uma á%
lgebra. Na tabela \ref{elalg} estão resumidas as propriedades dos elementos
de medida.

\begin{table}[h]
\begin{center}
\begin{tabular}{cl}
\textbf{Operação} & \textbf{Existência} \\ \hline
\multicolumn{1}{l}{} &  \\ 
\multicolumn{1}{l}{$\mathbf{1}\cdot M_{a}=M_{a}\cdot \mathbf{1}=M_{a}$} & 
Existe elemento identidade para o produto \\ 
\multicolumn{1}{l}{} &  \\ 
\multicolumn{1}{l}{$\mathbf{0}\cdot M_{a}=M_{a}\cdot \mathbf{0}=\mathbf{0}$}
& Existe elemento nulo para o produto \\ 
\multicolumn{1}{l}{} &  \\ 
\multicolumn{1}{l}{$M_{a}+\mathbf{0}=M_{a}$} & Existe elemento identidade
para a soma%
\end{tabular}%
\end{center}
\caption{Elementos da Álgebra de Medida}
\label{elalg}
\end{table}

\subsection{Medição de Observáveis Compatíveis}

Dois observáveis $A^{1}$\ e\ $A^{2}$ são ditos compatíveis se, filtrado do
sistema um estado associado ao valor $a_{1}^{1}$\ de $A^{1}$ uma filtragem
posterior de algum valor $a_{1}^{2}$\ pertencente ao observável $A^{2}$\ não
altera a informação adquirida no primeiro estágio da medição. Desta forma, o
sistema possuirá, com certeza, os valores $a_{1}^{1}$ e $a_{1}^{2}$ de uma
maneira bem definida e simultânea\footnote{%
Em $a_{k}^{j}$, o indice $k$ está relacionado com o lugar do elemento no
conjunto $E[A^{j}]=\{a_{k}^{j}\}=\{a_{1}^{j},a_{2}^{j},...\}$, com $%
k=\{1,...,N\}$; e o índice $j$, está relacionado com o observável, quando
ele pertence a uma família de características compatíveis $\mathbf{A}%
=\{A^{j}\}=\{A^{1},A^{2},...\}$, com $j=\{1,...,n\}$.}.

\begin{figure}[ht]
                           \begin{center}
                           \includegraphics[width=10 cm, height=4 cm]{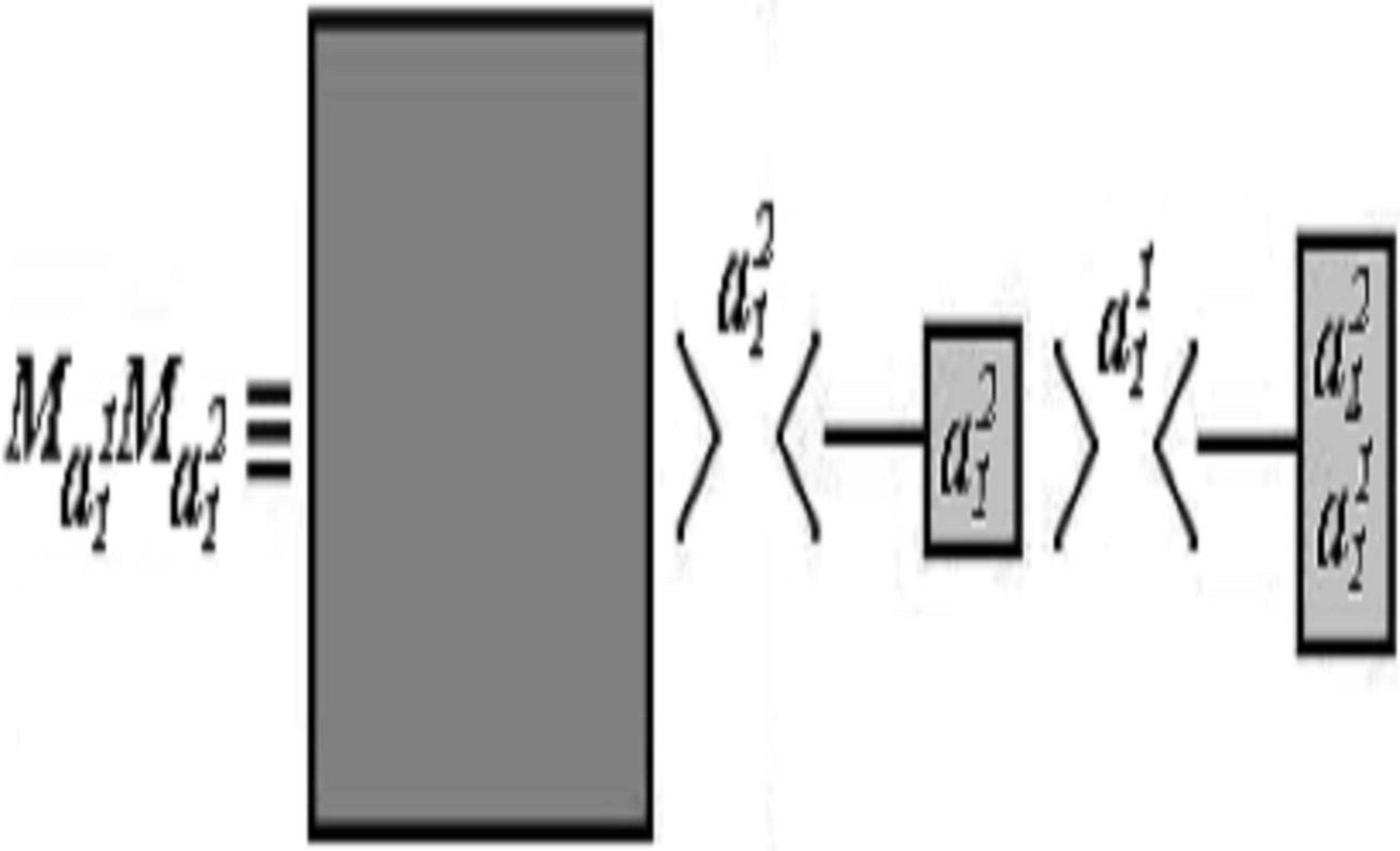}
                           \caption{O feixe ingressante ao segundo estágio do
processo tem o valor $a_{1}^{1}$ para o observ;avel $A^{1}$, a filtragem
posterior filtra o valor $a_{1}^{2}$ para $A^{2}$, e o sistema final tem os
valores $a_{1}^{1}$ e $a_{1}^{2}$.}
                           \end{center}
                           \label{fig6}
 \end{figure}

Um símbolo para uma operação como
esta pode ser construído da seguinte forma:%
\begin{equation}
M_{a_{1}^{2}a_{1}^{1}}=M_{a_{1}^{2}}M_{a_{1}^{1}}=M_{a_{1}^{1}}M_{a_{1}^{2}}.
\end{equation}%
O fato de que as medições efetuadas sejam operações compatíveis permite que
comutem, logo, temos%
\begin{equation}
M_{a^{1}a^{2}...a^{\max [j]}}=\prod_{k=1}^{\max [j]}M_{a^{k}}.
\end{equation}%
O maior conjunto de observáveis $\mathbf{A}=\left\{
A^{1},A^{2},A^{3},A^{4}...\right\} $ com que se pode caracterizar um sistema 
é chamado de \textbf{maximal}. \ Com este conjunto podemos ter o máximo
conhecimento do sistema. Assim, a medida de qualquer propriedade que não
pertença ao conjunto, ou que não possa ser expressa como uma combinação dos
elementos deste, vai alterar o estado do sistema mudando o conhecimento
ganho antes da última medição. \ Estes novos símbolos têm as mesmas
propriedades dos símbolos $M_{a_{k}}$\footnote{%
O delta de Kronecker para uma medida consecutiva de dois conjuntos de observá%
veis compatíveis é simbolizado por%
\begin{equation*}
\delta (a_{1}a_{2}a_{3}a_{4}a_{5}a_{6}...a_{\max [j]},a_{1}^{\prime
}a_{2}^{\prime }a_{3}^{\prime }a_{4}^{\prime }a_{5}^{\prime }a_{6}^{\prime
}...a_{\max [j]}^{\prime })=\prod_{k=1}^{\max [j]}\delta \left(
a_{k},a_{k}^{\prime }\right) 
\end{equation*}%
}.

A partir da próxima seção usaremos uma notação de $a$ para nos referir a um
elemento genérico $a_{j}\in \mathbf{E[}A\mathbf{]}$ associado ao observável $%
A$, dado que agora trataremos com operações genéricas entre os elementos de
diferentes conjuntos.

\subsection{Medição de Observáveis não Compatíveis}

Nesta seção estudaremos os sistemas que mudam seu estado ao fazermos medidas
deles, já que estes tipos de eventos descrevem de uma forma mais verossímil
os resultados de medidas reais. Desta forma, temos que estudar as medições
de observáveis não-compatíveis.

Dada a possibilidade de descrever um sistema por vários observáveis, temos
que se o sistema estiver descrito por um observável $A\Rightarrow \{\mathbf{E%
}[A],M_{a}\}$, este também poderia estar bem descrito por $B\Rightarrow \{%
\mathbf{E}[B],M_{b}\}$\footnote{%
Usamos a notação $A\Rightarrow \{\mathbf{E}[A],M_{a}\}$, para dizer que o
observável $A$ tem associado um espectro $\mathbf{E}[A]$ e um conjunto de sí%
mbolos de medida $M_{a}$.}.\ Assim, para começarmos esse estudo, devemos
analisar a possibilidade de construir um símbolo que represente as medidas
consecutivas de quantidades que pertençam a conjuntos de observáveis que não
são compatíveis entre si.\ Um processo como este envolve a medida
consecutiva de, pelo menos, dois observáveis, como a medida representada por 
$M_{a}M_{b}$. Se analisarmos este processo, vemos que pode ser representado
pelo seguinte esquema

\begin{figure}[ht]
                           \begin{center}
                           \includegraphics[width=10 cm, height=4 cm]{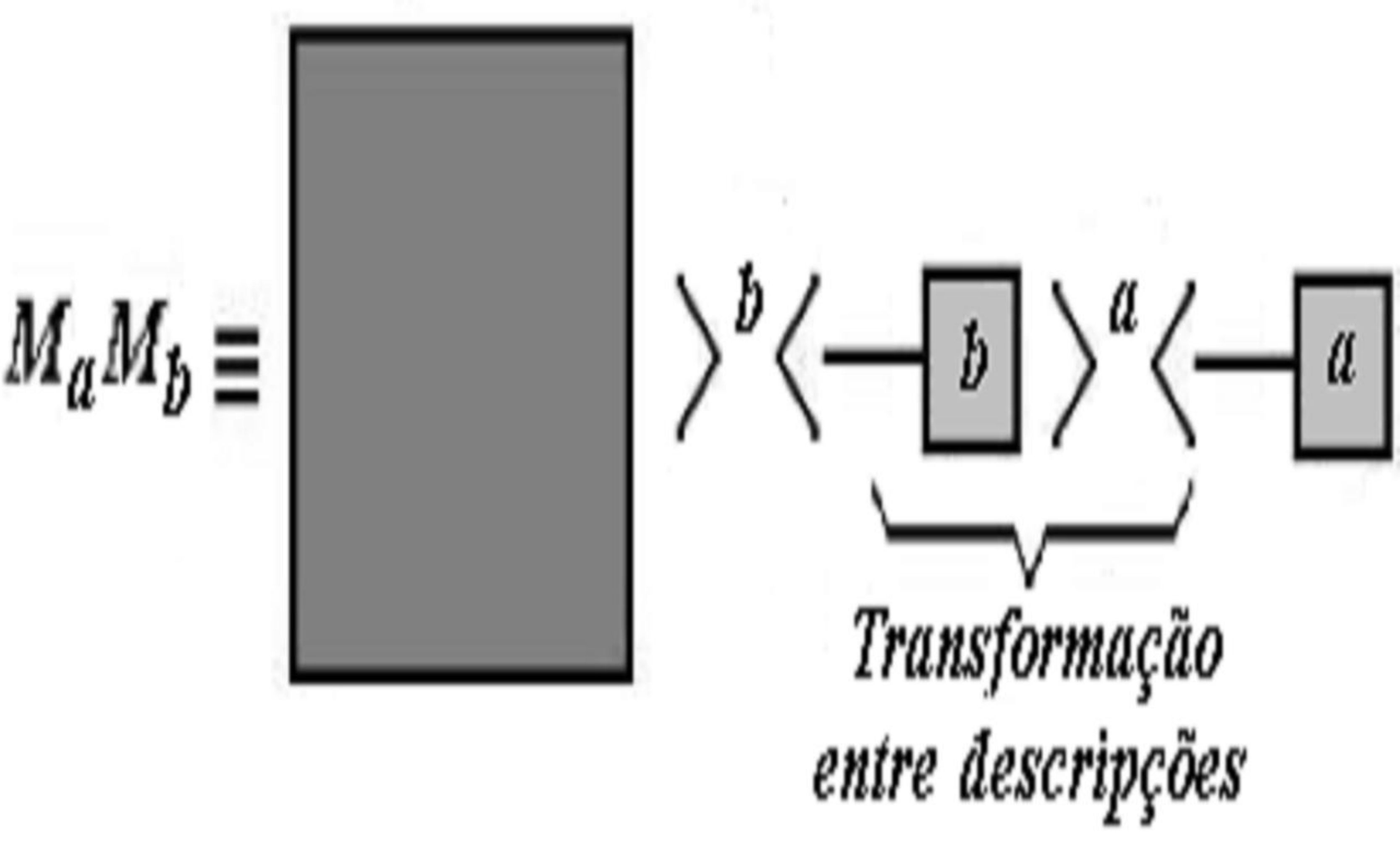}
                           \caption{Esta seqüência de medidas, em especial, escolhe
subsistemas que têm alguma propriedade $b$\ de $B$\ e, em seguida são
medidos no sistema (feixe) emergente aqueles subsistemas que tenham o valor $%
a$ de $A$.}
                           \end{center}
                           \label{noncompatible}
 \end{figure}

Dadas as características dos símbolos de medida até agora estudados, e o
fato que existe uma mudança na descrição do sistema, este processo não pode
ser representado por meio dos símbolos de medida até agora conhecidos. \
Portanto, nosso conjunto de relações tem que ser aumentado com um novo tipo
de símbolo.

Se analisarmos a representação do processo dada na figura \ref{noncompatible}%
, observamos que a mudança de representação envolve a existência de uma
escolha de um estado com a propriedade $a$ de $A$ entre aqueles emergentes
do estágio de medida que fez a escolha de $b$ de $B$.\ Tal símbolo pode ser 
\begin{equation}
M_{a}M_{b}=\left\langle a|b\right\rangle M_{a}^{b},  \label{med1}
\end{equation}%
onde o símbolo $M_{a}^{b}$\ vai representar o processo pelo qual um feixe
incidente em um estado de entrada $b$\ é mudado para um estado de saída $a$,
e o símbolo $\left\langle a|b\right\rangle $\ representa a \emph{transformaçã%
o} entre as duas representações $A$ e $B$\footnote{%
O processo em que não há nenhuma mudança pode ser simbolizado por $M_{a}^{a}$%
; este símbolo é equivalente ao processo $M_{a}$.}. Assim, para uma série de
medidas consecutivas 
\begin{equation}
M_{a}^{b}M_{c}^{d},
\end{equation}%
tomamos um feixe no estado $d$ de $D$ e o transformamos em um feixe no
estado $c$ de $C$. Em seguida é realizada outra medida em que só são aceitos
sistemas com o valor $b$\ de $B$\ entre aqueles estados que são originados
na saída do primeiro processo, e os mesmos são transformados em estados com
o valor bem definido $a$\ de $A$.

Para uma medida tal como a anterior, temos um resultado líquido de receber
um sistema num estado $d$ de $D$ e transformá-lo num estado $a$ de $A$. Como
cada símbolo de medida representa um estágio da medição, como resultado
final, temos uma mudança na representação do sistema dada pelos observáveis $%
B$ e $C$, e expressa pela quantidade $\left\langle b|c\right\rangle $.
Assim, podemos representar o produto destes símbolos da seguinte forma%
\begin{equation}
M_{a}^{b}M_{c}^{d}=\left\langle b|c\right\rangle M_{a}^{d}.
\end{equation}%
Se fizermos o processo na ordem contrária, temos que%
\begin{equation}
M_{c}^{d}M_{a}^{b}=\left\langle d|a\right\rangle M_{c}^{b},  \label{rela1}
\end{equation}%
o que nos mostra que os símbolos de medida para observáveis não-compatíveis n%
ão são comutativos, i.e., 
\begin{equation}
M_{a}^{b}M_{c}^{d}\neq M_{c}^{d}M_{a}^{b}.
\end{equation}

A mudança de representação, dada pelo símbolo $\left\langle a|b\right\rangle 
$, que é situada quando fazemos uma medida como em (\ref{med1}), implica que
alguns destes estados podem não passar pelo processo de filtragem, já que
nada garante que todos estejam neste estado. Isso difere do caso onde temos
uma medida que aceita ou rejeita todos os estado ingressantes, como pode ser
estabelecido se temos estados do mesmo observável,%
\begin{equation}
M_{a_{1}}M_{a_{2}}=\delta (a_{2},a_{1})M_{a_{2}}^{a_{1}}.
\end{equation}

Desta forma o complemento introduzido à simbologia de medida, $\left\langle
a|b\right\rangle $, expressa a possibilidade de uma medição não nula da
propriedade $a$ sobre o sistema emergente do processo de filtragem do
primeiro estágio de medição que mostrou o valor $b$. Este é o elemento que
contém a relação estatística, dado que se tem, de fato, uma espécie de
escolha entre os sistemas no estado $b$.

\section{Funções de Transformação}

Os elementos $\left\langle a|b\right\rangle $, associados à transformação
entre caracterizações, ou mudanças de representação, de um mesmo sistema quâ%
ntico, são de grande importância no entendimento dos efeitos das ações de
medida em sistemas microscópicos\footnote{%
Dado que esses elementos tem uma relação com a estatística de seleção de
estados, vamos tomá-los como pertencentes a um corpo de números que comutam
com os símbolos de medida.}. Esses elementos, que chamaremos de agora em
diante \emph{Funções de Transformação}, relacionam as medidas realizadas
sobre um sistema com a ajuda de dois observáveis, geralmente não compatíveis.%
\textrm{\ }Dado que cada símbolo de medida, seja $M_{a}=M_{a}^{a}$ ou $%
M_{a}^{b}$, representa um estágio ou unidade no processo de medição, temos
que pelas relações entre os símbolos de medida, como em (\ref{med1}), os
processos $M_{a}M_{b}^{c}$\ e $M_{a}^{b}M_{c}$, podem ser postos como um sí%
mbolo só, ou seja,%
\begin{equation}
M_{a}M_{b}^{c}=\left\langle a|b\right\rangle M_{a}^{c}\text{ \ \ e, \ \ }%
M_{a}^{b}M_{c}=\left\langle b|c\right\rangle M_{a}^{c},  \label{reltr}
\end{equation}%
onde temos em conta os estados inicial e final, combinado com a função de
transformação entre os dois processos.

\begin{figure}[ht]
                           \begin{center}
                           \includegraphics[width=15 cm, height=8 cm]{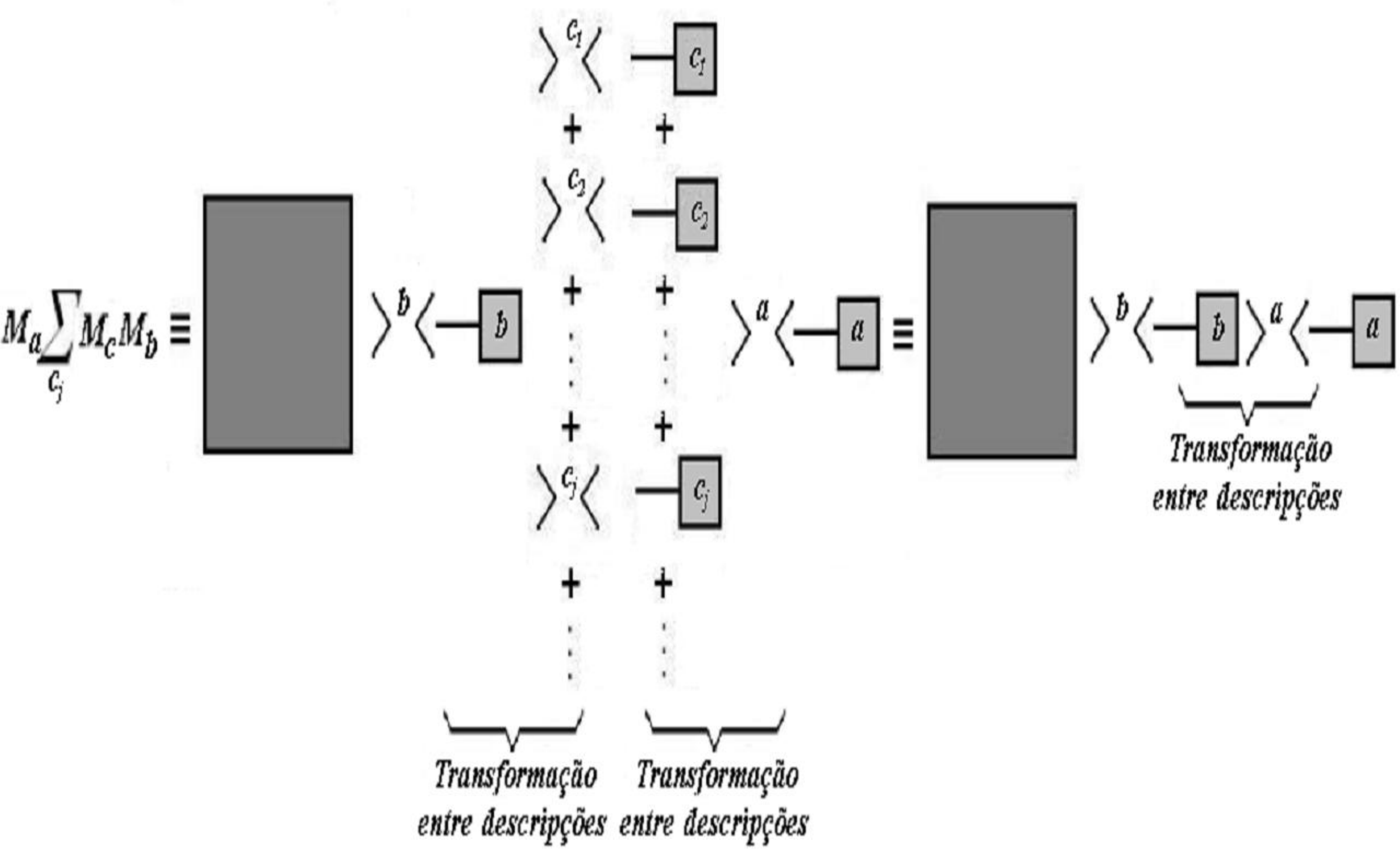}
                           \caption{Neste esquema, podemos entender como se dá a mudança de representação
de um sistema.}
                           \end{center}
                           \label{fig8}
 \end{figure}

Agora, se usarmos os conceitos de medida completa e as relações dadas em (%
\ref{reltr}) e (\ref{med1}), temos que, se realizarmos uma medida completa
na metade do processo que representa as medidas consecutivas $M_{a}M_{b}$,
podemos observar que%
\begin{equation}
M_{a}M_{b}=\left\langle a|b\right\rangle M_{a}^{b}=M_{a}\left(
\sum_{c}M_{c}\right) M_{b}=\sum_{c}\left\langle a|c\right\rangle
\left\langle c|b\right\rangle M_{a}^{b}.  \label{reltot}
\end{equation}

Se compararmos ambos resultados, podemos extrair que%
\begin{equation}
\left\langle a|b\right\rangle =\sum_{c}\left\langle a|c\right\rangle
\left\langle c|b\right\rangle .  \label{trans1}
\end{equation}%
Agora, se adotarmos a notação%
\begin{equation}
\sum_{c}\left\vert c\right\rangle \left\langle c\right\vert =\mathbf{1},
\label{sum1}
\end{equation}%
podemos inferir uma importante relação:%
\begin{equation}
M_{c}=\left\vert c\right\rangle \left\langle c\right\vert .  \label{simbolus}
\end{equation}

As considerações anteriores a respeito das transformações entre medidas també%
m nos permitem identificar a forma dos símbolos $M_{b}^{a}$\ da seguinte
maneira: se fizermos uma medida completa na saída do processo representado
pelo símbolo $M_{a}$, podemos ver que o resultado pode ser expresso como
combinação linear dos símbolos associados com $M_{b}$ da seguinte forma,%
\begin{equation}
\mathbf{1}M_{a}=\sum_{c}M_{c}M_{a}=\sum_{b}\left\langle b|a\right\rangle
M_{b}^{a},
\end{equation}%
ou, de forma equivalente, por meio de (\ref{simbolus}):%
\begin{equation}
\mathbf{1}M_{a}=\left( \sum_{b}\left\vert b\right\rangle \left\langle
b\right\vert \right) \left\vert a\right\rangle \left\langle a\right\vert .
\end{equation}

Assim, por comparação, podemos fazer a seguinte relação 
\begin{equation}
M_{b}^{a}=\left\vert b\right\rangle \left\langle a\right\vert .
\label{trocasimbolus}
\end{equation}%
Como consequência, temos que, se identificarmos $b=a^{\prime }$ em (\ref%
{reltot}), obtemos 
\begin{equation}
M_{a}M_{a^{\prime }}=\left\langle a|a^{\prime }\right\rangle
M_{a}^{a^{\prime }},
\end{equation}%
a qual será nula, se não tivermos $a^{\prime }=a$, o que nos mostra que 
\begin{equation}
\left\langle a|a^{\prime }\right\rangle =\delta (a,a^{\prime }),
\label{ortog}
\end{equation}%
que é o delta de Kronecker, obtido anteriormente em (\ref{delt}).

Como foi comentado na seção anterior, o fato que entre duas medições de dois
observáveis não compativeis possa passar parcial ou totalmente o feixe
incidente, mostra que funções de transformação podem ser interpretadas como
um fator que diz o quão compativeis são os observáveis. Além disto, também
pode-se inferir que em (\ref{rela1}) $\left\langle d|a\right\rangle $ está
relacionada com a possibilidade de que nos sistemas com valor $a$ de $A$
possam ser medidos estados com valor $d$ de $D$, o que está relacionado com
a interpretação estatística que será dada mais adiante.

\section{O Traço}

A função de transformação $\left\langle a|b\right\rangle $ pode ser
considerada como um funcional linear do símbolo de medida$\ M_{b}^{a}$, já
que ela está diretamente relacionada com as transformações das medidas
realizadas de quantidades não compatíveis. Tal correspondência é chamada 
\emph{traço} e tem a seguinte forma:%
\begin{equation}
\mathbf{Tr}\left\{ M_{b}^{a}\right\} =\mathbf{Tr}\left\{ \left\vert
b\right\rangle \left\langle a\right\vert \right\} =\left\langle
a|b\right\rangle ,  \label{trac}
\end{equation}%
esta forma específica vem do fato de que o corpo numérico trabalhado aqui
como escalares são os números complexos\footnote{%
Pode ser usado um outro corpo de números, como são os quaternions que podem
ser vistos em \cite{cassiuscu}, e \cite{quaternionscu}.}. Assim, se o
elemento de medida pertence à filtragem de um par de observaveis
compativeis, tem-se%
\begin{equation}
\mathbf{Tr}\left\{ M_{\overline{a}}^{a}\right\} =\left\langle a|\overline{a}%
\right\rangle =\delta \left( \overline{a},a\right) ,
\end{equation}%
e%
\begin{equation}
\mathbf{Tr}\left\{ M_{b}^{b}=M_{b}\right\} =1.
\end{equation}

Também podemos ver que o traço de um produto de símbolos de medida é dado por%
\begin{equation}
\mathbf{Tr}\left\{ M_{d}^{c}M_{b}^{a}\right\} =\mathbf{Tr}\left\{ \left\vert
d\right\rangle \left\langle c|b\right\rangle \left\langle a\right\vert
\right\} =\mathbf{Tr}\left\{ \left\langle c|b\right\rangle M_{d}^{a}\right\}
=\left\langle c|b\right\rangle \left\langle a|d\right\rangle ,  \label{trac1}
\end{equation}%
e o traço do produto invertido é dado por%
\begin{equation}
\mathbf{Tr}\left\{ M_{b}^{a}M_{d}^{c}\right\} =\mathbf{Tr}\left\{
\left\langle a|d\right\rangle M_{b}^{c}\right\} =\left\langle
a|d\right\rangle \left\langle c|b\right\rangle ,  \label{trac2}
\end{equation}%
desta forma, comparando (\ref{trac1}) com (\ref{trac2}), vemos que o traço
de um produto de símbolos de medida é comutativo, embora o produto dos sí%
mbolos não o seja.

\section{Interpretação Estatística}

Dadas as observações nas seções anteriores, temos claro que existe uma
importante componente estatística no sentido de como podemos interpretar o
resultado das medições sobre um sistema, e como essas deveriam ser
interpretadas. Um dos resultados mais relevantes, é que se temos duas descriç%
ões do mesmo sistema por meio dos estados $\{a_{i}\}\in \mathbf{E}[A]$ e $%
\{b_{j}\}\in \mathbf{E}[B]$ associados\ com dois observáveis diferentes $A$
e $B$, estas duas descrições podem ser relacionadas pelo conjunto de funções
de transformação $\{\langle a_{i}|b_{j}\rangle \}_{i,j}$. Se definimos dois
conjuntos de elementos $\left\{ \lambda (a)\right\} $ e $\left\{ \lambda
(b)\right\} $\ podemos ver que, se redefinimos os símbolos de medida como%
\begin{equation}
M_{b}^{a}\rightarrow \lambda (a)M_{b}^{a}\lambda ^{-1}(b)  \label{ie1}
\end{equation}%
e as funções de transformação como%
\begin{equation}
\left\langle a|b\right\rangle \rightarrow \lambda ^{-1}(a)\left\langle
a|b\right\rangle \lambda (b),  \label{ie2}
\end{equation}%
as operações entre os elementos da álgebra de medida não são alteradas. Dado
isto, temos que as funções de transformação $\left\langle a|b\right\rangle $%
\ não podem ter um significado físico direto, pelo fato de que não são
univocamente definidas, dada a arbitrariedade dos fatores $\lambda $.

A interpretação estatística se dá pelo argumento usado para o entendimento
do significado do símbolo $M_{b}^{a}M_{d}^{c}$. Como pode-se ver, no estágio
intermediário deste processo, no feixe com a propriedade $d$ do observável $%
D $ resultante do processo inicial $M_{d}^{c}$, se faz uma nova medida, mas
agora da propriedade $a$ do observável $A$, associada ao símbolo $M_{b}^{a}$%
. Assim, este fato pode ser tomado como um processo estatístico, dado que
temos um conjunto de sistemas num estado $d$ de $D$, há um ato de "\textbf{%
escolha}", ao medir entre eles o estado $a$ de $A$. Isto equivale a
interpretar o símbolo $\left\langle d|a\right\rangle $, como alguma
quantidade relacionada com a probabilidade de encontrar um sistema no estado 
$d$ quando temos um ensemble no estado $a$.

Para exemplificar um pouco mais a interpretação anterior, podemos
con\-si\-de\-rar o seguinte procedimento de medida,%
\begin{eqnarray}
M_{b}M_{a}M_{b} &=&\left\langle b|a\right\rangle M_{b}\left\langle
a|b\right\rangle \\
&=&\left\langle b|a\right\rangle \left\langle a|b\right\rangle M_{b},  \notag
\end{eqnarray}%
neste processo, aparentemente, não acontece nada, no sentido que são
recebidos sistemas com a propriedade $b$ de $B$ e os estados de saída têm,
em princípio, a mesma propriedade. Mas a passagem pela filtragem intermediá%
ria, a medição efetuada pelo símbolo $M_{a}$, origina a necessidade de que,
para obter algum resultado não nulo no estágio final da medida, as transforma%
ções entre os estados de $b\rightarrow a$ e, de novo, $a\rightarrow b$, têm
que ser possíveis, o que pode ser representado pelo símbolo%
\begin{equation}
p\left( a|b\right) =\left\langle b|a\right\rangle \left\langle
a|b\right\rangle ,
\end{equation}%
que, nesta primeira abordagem, refere-se à transformação sucessiva entre os
dois sistemas. Este novo símbolo é invariante frente à transformação (\ref%
{ie2}), dado que 
\begin{eqnarray}
p\left( a|b\right) &=&\left\langle b|a\right\rangle \left\langle
a|b\right\rangle =\lambda ^{-1}(b)\left\langle b|a\right\rangle \lambda
(a)\lambda ^{-1}(a)\left\langle a|b\right\rangle \lambda (b) \\
&=&\left\langle b|a\right\rangle \left\langle a|b\right\rangle ,  \notag
\end{eqnarray}%
o que faz deste um objeto que, além de estar univocamente definido, pode ser
relacionado diretamente com uma interpretação\footnote{%
Interpretação que é, basicamente, derivada da seleção do elemento dentre um
conjunto de resultados possíveis.} estatística como a \textit{probabilidade
de encontrar o sistema no estado }$a$\textit{\ quando é executada uma mediçã%
o sobre um sistema que se encontra no estado }$b$.

Se fizermos uma filtragem não-seletiva sobre todos os estados membros do
conjunto associado com o observável $A$, 
\begin{eqnarray}
M_{b}\hat{1}M_{b} &=&\sum_{a}M_{b}M_{a}M_{b}=\sum_{a}\left\langle
b|a\right\rangle \left\langle a|b\right\rangle M_{b} \\
&=&\sum_{a}p\left( a|b\right) M_{b},  \notag
\end{eqnarray}%
podemos encontrar que a quantidade $p\left( a|b\right) $, é normalizada à
unidade 
\begin{equation}
\sum_{a}p\left( a|b\right) =1.  \label{label}
\end{equation}%
Uma outra característica deste símbolo é a sua simetria, ou seja, a
pro\-ba\-bi\-li\-da\-de de que o sistema passe de $a\rightarrow b$, é a
mesma que $b\rightarrow a$, assim:%
\begin{equation}
p\left( a|b\right) =p\left( b|a\right) .
\end{equation}%
Além das propriedades encontradas para a probabilidade, temos outra apreciaçã%
o. Dado que os processos estudados anteriormente envolvem a transição entre
dois estados, a determinação de cada um destes estados envolve também dois
processos de medida. Desta maneira, existe a probabilidade de que somente
uma fração, todos ou nenhum dos estados envolvidos no primeiro estágio do
processo de medida sejam aceitos no segundo estágio. Isto significa que
existem dois valores extremos para a fração de estados que passa; que é $1$
se passar a totalidade ou, $0$ se não passar nenhum. Ou seja, a
probabilidade $p\left( a|b\right) $\ está entre os dois limites,%
\begin{equation*}
0\leq p\left( a|b\right) \leq 1.
\end{equation*}

Supondo que o número $\left\langle b|a\right\rangle $ é um número complexo
e, dado que a probabilidade é um número real maior que zero, podemos tomar $%
\left\langle b|a\right\rangle =\overline{\left\langle a|b\right\rangle }$, o
que garante que,%
\begin{equation}
p\left( a|b\right) =\overline{\left\langle a|b\right\rangle }\left\langle
a|b\right\rangle =\left\vert \left\langle a|b\right\rangle \right\vert
^{2}\geq 0.
\end{equation}

Um fato interessante desta escolha é a de que os conjuntos $\left\{ \lambda
(a)\right\} $ e $\left\{ \lambda (b)\right\} $ devem ser tais que a
probabilidade seja independente deles, assim temos que se 
\begin{equation}
\left\langle a|b\right\rangle =\lambda ^{-1}(a)\left\langle a|b\right\rangle
\lambda (b),  \label{a}
\end{equation}%
\begin{equation}
\overline{\left\langle a|b\right\rangle }=\overline{\lambda ^{-1}}(a)%
\overline{\left\langle a|b\right\rangle }\overline{\lambda }(b),  \label{b}
\end{equation}%
\begin{equation}
\left\langle b|a\right\rangle =\lambda ^{-1}(b)\left\langle b|a\right\rangle
\lambda (a),  \label{c}
\end{equation}%
multiplicando (\ref{a}) e (\ref{b}) e exigindo a invariância do resultado
obtemos%
\begin{eqnarray*}
\lambda (b)\overline{\lambda }(b) &=&1, \\
\lambda ^{-1}(a)\overline{\lambda ^{-1}}(a) &=&1,
\end{eqnarray*}%
o que significa que 
\begin{equation}
\overline{\lambda }(a)=\lambda ^{-1}(a).
\end{equation}%
Desta forma, podemos ver que pode-se associar uma representação exponencial
a estes números como uma fase da forma:%
\begin{equation}
\lambda (a)=e^{i\varphi (a)},
\end{equation}%
dependendo somente do elemento do espectro. O valor assumido pelas fases $%
\varphi (a)$ é arbitrário e não afeta o resultado das medições de uma forma
direta, já que as probabilidades independem desta fase.

\subsection{O Símbolo de Medida Adjunto}

O fato que%
\begin{equation}
p\left( a|b\right) =p\left( b|a\right) ,
\end{equation}%
envolve a existência de uma equivalência entre os processos representados
pelos símbolos de medida $M_{a}M_{b}$ e, seu processo inverso\footnote{%
Esta equivalência vem do fato que os eventos de seleção são \textbf{"estatí%
sticamente independentes"}. Para entender um pouco mais este conceito pocure
pelo teorema de Bayes e probabilidade condicional., \emph{Thomas Bayes- matem%
ático (Londres, Inglaterra, 1702 - Tunbridge Wells, 1761)}.}, o símbolo $%
M_{b}M_{a}$. Podemos ver que, tomando (\ref{med1})%
\begin{equation}
M_{a}M_{b}=\left\langle a|b\right\rangle M_{a}^{b}\text{ \ \ e, \ \ }%
M_{b}M_{a}=\left\langle b|a\right\rangle M_{b}^{a},
\end{equation}%
e estabelecendo a conexão entre os os processos anteriormente representados
com a seguinte convenção%
\begin{equation}
\left( M_{a}M_{b}\right) ^{\dag }=M_{b}M_{a},
\end{equation}%
onde o símbolo $\dag $ significa\footnote{%
Lê-se o símbolo $\dagger $ como \emph{dagger}, termo anglosaxônico que
significa \emph{adaga}.} a operação adjunta. Temos que%
\begin{equation}
\left( M_{a}M_{b}\right) ^{\dag }=\overline{\left\langle a|b\right\rangle }%
\left( M_{a}^{b}\right) ^{\dag }=\overline{\left\langle a|b\right\rangle }%
M_{b}^{a}=M_{b}M_{a}=\left\langle b|a\right\rangle M_{b}^{a},
\end{equation}%
nos mostra que\footnote{%
As funções de transformação, $\left\langle a|b\right\rangle $, são números
complexos, e a linha $\overline{\hspace{0.3cm}}$ sobre estes números é
interpretada aqui como complexo conjugado.}%
\begin{equation}
\overline{\left\langle a|b\right\rangle }=\left\langle b|a\right\rangle .
\end{equation}%
Temos como um caso especial%
\begin{equation}
\left( M_{a}M_{a^{\prime }}\right) ^{\dag }=M_{a^{\prime }}M_{a},
\end{equation}%
onde pela igualdade $\delta (a^{\prime },a)=\delta (a,a^{\prime })$\ temos a
particularidade que%
\begin{equation}
M_{a}^{\dag }=M_{a},
\end{equation}%
o que define este símbolo como auto-adjunto, ou seja igual a seu adjunto.

Para a soma, e demais operações entre os símbolos de medida pode-se resumir
em%
\begin{eqnarray}
\left( X+Y\right) ^{\dag } &=&X^{\dag }+Y^{\dag }, \\
\left( X\cdot Y\right) ^{\dag } &=&Y^{\dag }\cdot X^{\dag }, \\
\left( \lambda \cdot Y\right) ^{\dag } &=&\overline{\lambda }\cdot Y^{\dag }.
\end{eqnarray}

\subsection{Álgebra Conjugada}

O uso dos números complexos como escalares na álgebra de medida implica na
existência de uma álgebra conjugada, uma transformação, na qual todos os nú%
meros são trocados pelos seus complexos conjugados. Assim,%
\begin{eqnarray}
\overline{\left( X+Y\right) } &=&\overline{X}+\overline{Y}, \\
\overline{\left( X\cdot Y\right) } &=&\overline{X}\cdot \overline{Y}, \\
\overline{\left( \lambda \cdot Y\right) } &=&\overline{\lambda }\cdot 
\overline{Y}.
\end{eqnarray}

A formação do adjunto dentro da álgebra conjugada dos símbolos de medida tem
a forma geral de%
\begin{equation*}
X^{T}=\overline{X}^{\dag }=\overline{X^{\dag }},
\end{equation*}%
e recebe o nome de \textit{transposição, }possuindo as propriedades:%
\begin{equation*}
\begin{array}{ccc}
\left( X+Y\right) ^{T}=X^{T}+Y^{T} & \left( XY\right) ^{T}=Y^{T}X^{T} & 
\left( \lambda Y\right) ^{T}=\lambda Y^{T}.%
\end{array}%
\end{equation*}

Essa operação deve ser diferenciada da operação de achar o adjunto, pois
esta tem o significado físico da inversão do processo representado pelo
operador.

\section{Representação Matricial de um Operador}

Como foi tratado nas primeiras seções, os símbolos de medida%
\begin{equation}
M_{a}=\left\vert a\right\rangle \left\langle a\right\vert \text{ \ \ \ e, \
\ }M_{b}^{a}=\left\vert b\right\rangle \left\langle a\right\vert ,
\end{equation}%
extraem informação de um sistema, originando um outro sistema, com caracterí%
sticas específicas; estes símbolos, junto com os observáveis $A$ e $B$, são
considerados na linguagem matemática como \textbf{operadores}, que são
entidades que agem sobre determinados objetos extraindo informação deles.
Estes operadores, assim como nossos observáveis e símbolos de medida, podem
ser representados equivalentemente. Como vimos na relação (\ref{reltot}), um
símbolo de medida de uma classe pode ser posto em função dos símbolos de
medida associados com um outro observável. Assim, podemos ver que, se
tivermos o símbolo de medida $M_{a}$, podemos expressá-lo em função dos sí%
mbolos de medida associados com $B$, fazendo uma medida completa da seguinte
forma%
\begin{equation}
M_{a}=\sum_{b}M_{b}M_{a}=\sum_{b}\left\langle b|a\right\rangle M_{b}^{a}.
\end{equation}

Conseqüentemente, se tomarmos de uma forma geral um observável $X$, por
exemplo, podemos expressá-lo pela sua influência sobre sistemas
caracterizados por observáveis conhecidos. Assim, se medirmos $X$ sobre um
sistema que está totalmente caracterizado por $A\Rightarrow \{M_{a},E[A]\}$,
e depois colocarmos na saída da operação representada por $X$ uma medida
completa, digamos, de $B\Rightarrow \{M_{b},E[B]\}$, teríamos $X$ em uma
representação mista dada por elementos de $A$ e $B$, da seguinte forma 
\begin{eqnarray}
X &=&\sum_{b}\sum_{a}M_{b}XM_{a}=\sum_{b}\sum_{a}\left\vert b\right\rangle
\left\langle b\right\vert X\left\vert a\right\rangle \left\langle
a\right\vert  \label{reprex} \\
&=&\sum_{a,b}\left\langle b\right\vert X\left\vert a\right\rangle M_{b}^{a},
\notag
\end{eqnarray}%
o que nos mostra a possibilidade de expressar qualquer operador por meio da
sua influência sobre um sistema conhecido.

Uma das características interessantes deste tipo de representação, é que
quando os conjuntos $\mathbf{E}[A]$ e $\mathbf{E}[B]$, são discretos, dada a
estrutura dos produtos entre operadores, os elementos $\left\langle
a\right\vert X\left\vert b\right\rangle $, em (\ref{reprex}) podem ser
considerados como os elementos de uma representação matricial do operador $X$%
, numa base mista de um espaço vetorial\footnote{%
Este raciocínio é uma das primeiras indicações a respeito de uma possível
associação com espaços vetoriais da álgebra de medida.}.

A forma (\ref{reprex}) permite expressar o produto de dois operadores como%
\begin{equation*}
XY=\sum_{a}\sum_{d}\left\langle a\right\vert XY\left\vert d\right\rangle
M_{a}^{d},
\end{equation*}%
de onde é derivado%
\begin{equation*}
\left\langle a\right\vert XY\left\vert d\right\rangle =\sum_{b}\left\langle
a\right\vert X\left\vert b\right\rangle \left\langle b\right\vert
Y\left\vert d\right\rangle .
\end{equation*}%
Os elementos da representação matricial de $X$ podem ser expressos de uma
forma e\-qui\-va\-len\-te pelo funcional\footnote{%
A palavra funcional se aplica aqui ao objeto que associa operadores a
escalares.} traço%
\begin{equation*}
\mathbf{Tr}\{M_{a}^{b}X\}=\left\langle a\right\vert X\left\vert
b\right\rangle ,
\end{equation*}%
onde podemos observar que, lembrando a propriedade (\ref{trac}) e a
linearidade do funcional, temos o seguinte%
\begin{equation*}
\mathbf{Tr\{}M_{a}^{b}X\mathbf{\}}=\mathbf{Tr}\left\{
\sum_{c}\sum_{d}\left\langle c\right\vert X\left\vert a\right\rangle
\left\langle b\right\vert \left. d\right\rangle M_{c}^{d}\right\}
=\sum_{c}\left\langle b\right\vert \left. c\right\rangle \left\langle
c\right\vert X\left\vert a\right\rangle =\left\langle a\right\vert
X\left\vert b\right\rangle ,
\end{equation*}%
em que um caso particular é dado por%
\begin{equation*}
\mathbf{Tr}\left\{ M_{a}X\right\} =\left\langle a\right\vert X\left\vert
a\right\rangle .
\end{equation*}%
De fato, o nome traço vem da relação:%
\begin{eqnarray*}
\mathbf{Tr}\left\{ X\right\} &=&\mathbf{\mathbf{Tr}}\left\{ \mathbf{1}\cdot
X\right\} \\
&\mathbf{=}&\mathbf{\mathbf{Tr}}\left\{ \sum_{c}M_{c}\cdot X\right\} \mathbf{%
=}\sum_{c}\left\langle c\right\vert X\left\vert c\right\rangle .
\end{eqnarray*}

As matrizes que representam os operadores adjuntos são matrizes conjugadas
complexas e transpostas das matrizes que representam os operadores
originalmente. Assim, o operador adjunto associado a $X$ é dado por%
\begin{equation*}
\left\langle a\right\vert X^{\dag }\left\vert b\right\rangle =\overline{%
\left\langle b\right\vert X\left\vert a\right\rangle }.
\end{equation*}%
Como casos especiais das representações anteriores, temos que se $X=1$, em $%
\mathbf{Tr}{\mathbf{\{}M_{a}^{b}X\mathbf{\}}}$, obtemos%
\begin{equation*}
\mathbf{Tr}\{\left( X=1\right) M_{a}^{b}\}=\left\langle a\right\vert \left(
X=1\right) \left\vert b\right\rangle =\left\langle a|b\right\rangle .
\end{equation*}

\section{Valor Esperado}

Trataremos do valor esperado de uma propriedade $A$ o que é obtido
multiplicando um valor particular $a$ pela probabilidade de obtê-lo, uma vez
que o estado se encontra inicialmente no estado $b$ de $B$. Assim, a
probabilidade de obter um valor específico $a\in \mathbf{E}[A]$, quando o
sistema está no estado $b$ aportará um valor específico à soma ponderada dos
elementos de $\mathbf{E}[A]$. A expressão matemática para o valor esperado $%
\left\langle A\right\rangle _{b}=\left\langle b\right\vert A\left\vert
b\right\rangle $ é dada por:%
\begin{equation}
\left\langle A\right\rangle _{b}=\sum_{a}p\left( a|b\right)
a=\sum_{a}\left\langle b|a\right\rangle a\left\langle a|b\right\rangle .
\end{equation}%
Usando a equação (\ref{trac2}), temos que a expressão anterior pode ser
posta como 
\begin{equation*}
\left\langle A\right\rangle _{b}=\sum_{a}\left\langle b|a\right\rangle
a\left\langle a|b\right\rangle =\mathbf{Tr}\left\{
M_{b}\sum_{a}aM_{a}\right\} .
\end{equation*}%
Podemos então definir a representação espectral do operador $A$ como $%
A=\sum_{a}a$\.{$\left\vert a\right\rangle \left\langle a\right\vert $, }de
tal forma que o valor esperado $\left\langle A\right\rangle _{b}$ também
pode ser escrito como%
\begin{equation*}
\left\langle A\right\rangle _{b}=\mathbf{Tr}{\left\{ AM_{b}\right\} .}
\end{equation*}

\section{A geometria dos estados}

A caracterização de um sistema microscópico por um conjunto $A\Rightarrow
\{M_{a},\mathbf{E}[A]\}$, nos mostra que cada símbolo de medida $%
M_{a}=\left\vert a\right\rangle \left\langle a\right\vert $ nos dá informaçã%
o sobre a existência de um de\-ter\-mi\-na\-do estado em um sistema. Como
podemos ver, o fato de que estes símbolos representam a extração de um
sub-sistema com uma característica única, e que o conjunto completo destes
sub-sistemas nos permite expressar outros estados asociados com outros observ%
áveis, nos leva a pensar os \emph{símbolos de medida} como \textit{projetores%
} sobre elementos de um espaço vetorial complexo, fato que torna possível a
associação com uma estrutura geométrica.

\subsection{Decomposição de Uma Medida}

Como foi dito anteriormente, a sensibilidade de um sistema quântico à medida
torna impossível seguir um estado durante o processo de medida. Desta forma,
por exemplo, no processo de medida representado por%
\begin{equation}
M_{b}M_{a}=\left\langle b|a\right\rangle M_{b}^{a}=\left\langle
b|a\right\rangle \left\vert b\right\rangle \left\langle a\right\vert ,
\label{med}
\end{equation}%
o co\-nhe\-ci\-men\-to dos estados intermediários entre $a$ e $b$ não é um
conhecimento útil no sentido estrito dos resultados físicos, já que somente
importam os estados \textit{iniciais }($\rightarrow a$) e \textit{finais }($%
b\rightarrow $). Isto, faz com que o processo representado por $%
M_{b}^{a}=\left\vert b\right\rangle \left\langle a\right\vert $, a mudança
do estado com valor $a$ de $A$ para um estado com valor $b$ de $B$, possa
ser considerado como um bloco indivisível.

Apesar disto, à parte $\left\vert b\right\rangle \left\langle a\right\vert $
do processo (\ref{med}), podemos dar uma interpretação um pouco mais
abstrata, em que são aceitos todos os sistemas que têm a propriedade $a$ de $%
A$, o sistema ingressante com esse estado quântico é "destruído" e depois é
"criado" um sistema no estado $b$ de $B$. Visto dessa forma, o processo de
medida pode ser dividido em duas partes: a destruição de um estado e a criaçã%
o de outro.

Assim, o símbolo $M_{b}^{a}$ pode ser visto como equivalente ao produto de
dois símbolos, $\left\langle a\right\vert $ que representa a destruição da
informação do sistema que tem a propriedade $a$ de $A$ e, $\left\vert
b\right\rangle $ que representa a criação de um sistema com a propriedade $b$
de $B$.

Dada esta nova maneira de interpretar os símbolos de medida, e lembrando a
relação obtida em (\ref{ortog}), podemos ver que a relação entre elementos
de criação e aniquilação de informação no sistema tende a se parecer com uma
relação de ortogonalidade entre vetores de um espaço vetorial 
\begin{equation*}
\overrightarrow{a}_{i}\cdot \overrightarrow{a}_{j}=\delta _{i,j}.
\end{equation*}%
Assim, sem perda de generalidade, podemos considerar os elementos $%
\left\vert a\right\rangle $ e $\left\langle b\right\vert $ como vetores de
um espaço vetorial complexo e o seu dual, respectivamente.

Do raciocínio anterior, podemos ver que todas as propriedades dos símbolos
de medida não mudam sob essa nova interpretação geométrica.

\subsection{Álgebra Vetorial}

Os elementos $\left\vert a\right\rangle $ e $\left\langle b\right\vert $,
agora considerados como vetores de um espaço vetorial complexo, podem também
ser considerados como uma representação do estado que criam ou destroem.
Assim como os símbolos de medida, estes elementos podem ser representados em
função dos resultados da medida de um outro observável, por meio de uma
medida completa, mas agora devemos considerar os elementos $\left\vert \text{
\ \ }\right\rangle \left\langle \text{ \ \ }\right\vert $ como os projetores
sobre um espaço vetorial associado com o observável que representam. Desta
forma, tomando (\ref{sum1}), podemos ver que o estado quântico $\left\vert
a\right\rangle $, do mesmo modo que seu dual, pode ser representado na base $%
\left\{ \left\vert b_{j}\right\rangle \right\} $, por%
\begin{equation}
\left\vert a\right\rangle =\sum_{j}\left\vert b_{j}\right\rangle
\left\langle b_{j}|a\right\rangle \text{.}  \label{combiaforb}
\end{equation}%
Aqui se faz uso das funções de transformação, já que a projeção do estado $%
\left\vert a\right\rangle $ sobre cada estado associado com o observável $B$
implica uma mudança de representação e, portanto, é equivalente a ver um
vetor de um espaço vetorial numa outra base, onde os elementos do vetor são
os números complexos dados por $\left\langle b|a\right\rangle $.

Isso nos leva a perceber que o elemento $\left\langle a|b\right\rangle $,
pode ser visto como um produto interno entre os elementos do espaço vetorial
sobre os números complexos. Dado que o produto interno está relacionado
diretamente com a geo\-me\-tria do espaço \cite{courant} , mais precisamente
com a norma, o elemento $\left\langle a|b\right\rangle $ induzirá caracterí%
sticas geométricas que podem ajudar a aplicar esta nova álgebra.

Desta maneira, as relações de linearidade impostas pela superposição nas
formas em que são decompostos os estados estabelecem, junto com as operações
já mencionadas, (soma, multiplicação por um escalar e mais adiante a norma),
o que é chamado uma \textbf{"álgebra de estados"}.

\subsubsection{Ação dos Operadores Sobre os Estados}

A ação de medida sobre um sistema, representada pela ação de um operador
sobre um objeto abstrato, que simboliza um estado, pode ser agora realizada
de uma forma que nos permite a comparação com procedimentos convencionais da 
álgebra linear. Como podemos ver, se tivermos um sistema representado por um
estado $\left\vert b\right\rangle $ e efetivarmos uma medida de um obervável 
$A$ sobre esse sistema, teremos que traduzir a informação que temos em $%
\left\vert b\right\rangle $ na linguagem do operador $A$. \ Assim, o
primeiro passo será representar o vetor $\left\vert b\right\rangle $ como
uma combinação linear dos elementos de $A$, segundo (\ref{combiaforb}) temos 
\begin{equation}
\left\vert b\right\rangle =\sum_{j}\left\vert a_{j}\right\rangle
\left\langle a_{j}|b\right\rangle .
\end{equation}

Como um segundo passo, podemos fazer a medida do observável $A$ sobre este
estado. Para este propósito, usamos a forma espectral desse operador, $%
A=\sum_{i}a_{i}$\.{$\left\vert a_{i}\right\rangle \left\langle
a_{i}\right\vert $},%
\begin{eqnarray*}
A\left\vert b\right\rangle &=&\sum_{i,j}a_{i}\left\vert a_{i}\right\rangle
\left\langle a_{i}|a_{j}\right\rangle \left\langle a_{j}|b\right\rangle \\
&=&\sum_{j}a_{j}\left\vert a_{j}\right\rangle \left\langle
a_{j}|b\right\rangle ,
\end{eqnarray*}%
onde foi usada a relação (\ref{ortog}). O resultado anterior, nos permite
observar um fato interessante: se o estado $\left\vert b\right\rangle
=\left\vert a_{i}\right\rangle $ for um estado associado com o operador $A$,
teríamos que 
\begin{equation}
A\left\vert a_{j}\right\rangle =\sum_{i,j}a_{i}\left\vert a_{i}\right\rangle
\left\langle a_{i}|a_{j}\right\rangle =a_{j}\left\vert a_{j}\right\rangle ,
\label{eigenek}
\end{equation}%
que é equivalente, na linguagem da álgebra linear, a uma \textit{equação de
valores-próprios}. Os elementos $\left\vert a_{i}\right\rangle $ são
elementos de um espaço vetorial que representam estados, e por isso são
chamados de \textit{vetores de estado}, e o conjunto de vetores $\left\{
\left\vert a_{i}\right\rangle \right\} _{j}$\ são chamados de conjunto de
vetores associados ao observável $A$, ou, simplesmente, conjunto de \textit{%
vetores-próprios} de $A$. O conjunto $E[A]$, será o conjunto de \textit{%
valores-próprios} de $A$, ou espectro de $A$.

A representação espectral de um operador, nos permite conhecer como sería a
representação espectral de uma função deste. Assim, se tivermos, por
exemplo, $A^{2}$, usando (\ref{eigenek}) encontramos 
\begin{equation*}
A^{2}=A\sum_{a}\left\vert a\right\rangle a\left\langle a\right\vert
=\sum_{a}\left( A\left\vert a\right\rangle \right) a\left\langle
a\right\vert =\sum_{a}\left\vert a\right\rangle a^{2}\left\langle
a\right\vert .
\end{equation*}%
Portanto, temos que se a função $f\left( A\right) $\ puder ser expressa numa
série de potências, podemos obter%
\begin{equation*}
f\left( A\right) =\sum_{a}\left\vert a\right\rangle f\left( a\right)
\left\langle a\right\vert .
\end{equation*}%
Com a representação espectral de um operador, podemos construir um polinô%
mio, cujas raízes serão o conjunto de valores próprios, da seguinte forma:
dado um elemento $a_{1}\in \mathbf{E}\left[ A\right] $, podemos mostrar que%
\begin{equation*}
A-a_{1}1=\sum_{a}\left\vert a\right\rangle \left( a-a_{1}\right)
\left\langle a\right\vert ,
\end{equation*}%
de tal forma que se tomarmos todos os pontos do conjunto $\mathbf{E}\left[ A%
\right] $, temos%
\begin{equation*}
\prod_{k}\left( A-a_{k}\right) =\sum_{a}\left\vert a\right\rangle
\prod_{k}\left( a-a_{k}\right) \left\langle a\right\vert .
\end{equation*}

A expressão anterior define um polinômio em $a$ 
\begin{equation*}
\prod_{k}\left( a-a_{k}\right) ,
\end{equation*}%
que tem suas raízes em cada ponto do conjunto $\mathbf{E}\left[ A\right] $.
Esta equação é chamada \textbf{polinômio característico}.

\subsection{Funções de Onda}

O espaço vetorial construído para descrever um sistema quântico qualquer,
fornece uma base para representar todo estado em que possa estar o sistema.
As propriedades do vetor, que representa o sistema, serão expressas no
conjunto de funções de transformação associadas com a projeção desse vetor
em cada um dos elementos da base do espaço vetorial. Esse conjunto de nú%
meros é conhecido como \emph{função de onda}. Os vetores de estado, possuem $%
N$ números (componentes, dependendo da dimensionalidade do sistema),
associados com os $N$ elementos da base.

Suponhamos que temos dois sistemas representados pelos vetores de estado $%
\left\vert \psi \right\rangle $ e $\left\vert \phi \right\rangle $, como
temos visto nas seções anteriores, podemos escreve-los na base $\left\{
\left\vert b\right\rangle \right\} $, como%
\begin{equation*}
\left\vert \psi \right\rangle =\sum_{b}\left\vert b\right\rangle
\left\langle b|\psi \right\rangle =\sum_{b}\left\vert b\right\rangle \psi
\left( b\right) ,
\end{equation*}%
e 
\begin{equation*}
\left\langle \phi \right\vert =\sum_{b}\left\langle \phi |b\right\rangle
\left\langle b\right\vert =\sum_{b}\phi ^{\dag }\left( b\right) \left\langle
b\right\vert ,
\end{equation*}%
onde%
\begin{equation*}
\phi ^{\dag }\left( b\right) =\left\langle \phi |b\right\rangle .
\end{equation*}%
Se $\psi $ e $\phi $ estão em relação adjunta, $\phi =\psi ^{\dag }$, a
correspondente função de onda está conectada pela ação de encontrar o
adjunto como%
\begin{equation*}
\phi \left( b\right) =\psi ^{\dag }\left( b\right) .
\end{equation*}

O produto de dois vetores de estado será dado por%
\begin{eqnarray*}
\left\langle \psi _{2}|\phi _{1}\right\rangle &=&\sum_{b}\left\langle \psi
_{2}|b\right\rangle \left\langle b|\phi _{1}\right\rangle \\
&=&\sum_{b}\psi _{2}^{\dag }\left( b\right) \phi _{1}\left( b\right) ,
\end{eqnarray*}%
e, em particular, 
\begin{equation*}
\left\langle \psi |\psi \right\rangle =\sum_{b}\psi ^{\dag }\left( b\right)
\psi \left( b\right) \geq 0.
\end{equation*}

A geometria dos estados é uma geometria unitária, já que a norma da função
de onda é invariante, ou seja: 
\begin{equation*}
\sum_{b}\psi ^{\dag }\left( b\right) \psi \left( b\right) =\sum_{a}\psi
^{\dag }\left( a\right) \psi \left( a\right) ,
\end{equation*}%
onde se tem que 
\begin{equation*}
\psi \left( a\right) =\sum_{a}U_{ab}\psi \left( b\right) ,
\end{equation*}%
e $U_{ab}$ é o operador unitário que a realiza a transformação entre uma
base e a outra.

Para especificar a forma explícita de $U_{ab}$, comecemos lembrando que o
operador $\left\vert \psi _{1}\right\rangle \left\langle \phi
_{2}\right\vert $ é representado pela matriz%
\begin{equation*}
\left\langle b|\psi _{1}\right\rangle \left\langle \phi _{2}|b\right\rangle
=\psi _{1}\left( b\right) \phi _{2}\left( b\right) ,
\end{equation*}%
e funções de onda que representam $X\left\vert \psi \right\rangle $ e $%
\left\langle \phi \right\vert X$ são%
\begin{equation*}
\left\langle a\right\vert X\left\vert \psi \right\rangle
=\sum_{b}\left\langle a\right\vert X\left\vert b\right\rangle \psi \left(
b\right) ,
\end{equation*}%
e%
\begin{equation*}
\left\langle \phi \right\vert X\left\vert b\right\rangle
=\sum_{a}\left\langle a\right\vert X\left\vert b\right\rangle \phi \left(
a\right) .
\end{equation*}%
Substituindo nas expressões anteriores o operador $X=1$, obtemos a relação
entre as funções de onda de um vetor em duas diferentes representações,%
\begin{eqnarray*}
\psi \left( a\right) &=&\sum_{b}\left\langle a|b\right\rangle \psi \left(
b\right) , \\
\phi \left( a\right) &=&\sum_{b}\phi \left( b\right) \left\langle
a|b\right\rangle ,
\end{eqnarray*}%
o que nos permite concluir que%
\begin{equation*}
U_{ab}=\left\langle a|b\right\rangle .
\end{equation*}

Nota-se que a função de onda que representa $\left\vert b\right\rangle $ na
descrição da base $\left\{ a\right\} $ é%
\begin{equation*}
\psi _{b}\left( a\right) =\left\langle a|b\right\rangle =\phi _{a}^{\dagger
}\left( b\right) .
\end{equation*}%
Do ponto de vista da álgebra de medida, as funções de onda $\phi ^{\dagger }$
e $\psi $ são matrizes com uma só linha ou coluna cada uma. \ Dado que
qualquer operador Hermiteano simboliza uma quantidade física, temos que
qualquer observável pode ser representado por um operador Hermiteano, e que
qualquer vetor unitário simboliza um estado. Então o valor esperado da
propriedade $X$ no estado $\left\vert \psi \right\rangle $ é dado por%
\begin{equation*}
\left\langle X\right\rangle _{\psi }=\left\langle \psi \right\vert
X\left\vert \psi \right\rangle =\sum_{a,b}\psi ^{\dag }\left( a\right)
\left\langle a\right\vert X\left\vert b\right\rangle \psi \left( b\right) .
\end{equation*}%
Em particular, a probabilidade de se observar o valor $a$ em uma medida
relacionada com $A$ feita sobre o sistema no estado $\left\vert \psi
\right\rangle $, é simbolizada por%
\begin{equation*}
p\left( a|\psi \right) =\left\langle \psi \left\vert a\right\rangle
\left\langle a\right\vert \psi \right\rangle =\psi ^{\dag }\left( a\right)
\psi \left( a\right) =\left\vert \psi \left( a\right) \right\vert ^{2},
\end{equation*}%
que é a nossa definição de probabilidade.

\section{Conclusões}

A visão de Schwinger dos processos de medida em M.Q. permite a introdução da
símbologia da medida que, em uma forma resumida, é compatível com as formas
conhecidas mais simples de interação aparelho-sistema. Esta simbologia
permite estabelecer relações básicas entre tais processos e a construção de
estruturas consistentes entre eles. Como pudemos ver ao longo deste artigo,
a simplicidade desta abordagem baseia-se no fato de que é necessário apenas
o conhecimento de conceitos procedentes da Álgebra Linear. Desta forma, a
construção de um espaço vetorial que está relacionado com a estrutura matemá%
tica e caracterização de um sistema quântico, nos resulta intuitiva. A import%
ância das transformações unitárias para as mudanças entre as diferentes
caracterizações de um sistema quântico também se destaca de maneira natural
dentro desta abordagem.

Na construção de um espaço vetorial para um sistema ao nível quântico,
primeiramente devemos ter em conta que não conhecemos nada do sistema até
realizarmos alguns processos de medição. Estes processos nos ajudam a
encontrar estados que são associados aos observáveis nos quais temos
interesse. Tais observáveis podem tomar uma determinada quantidade de
valores que são associados ao conjunto que chamamos de espectro, e cada
sistema que tem algum desses valores bem definido, é dito estar nesse estado
(Exemplo: o sistema que tem o valor bem definido $\underline{a}$\ da
quantidade $A$, se diz estar no estado $\left\vert \underline{a}%
\right\rangle $), que é representado por um elemento de um espaço vetorial
complexo como $\left\vert a\right\rangle $. Este espaço vetorial possui
dimensão $N$, com $N$ o número de estados possíveis do observável com que
caracterizamos o sistema.

Portanto, na caracterização de um sistema por meio das medições de um observá%
vel, associamos um espaço vetorial complexo (chamado de \emph{espaço de
Hilbert}), normado, com uma estrutura de medida bem definida. O conjunto de
estados que constituem o espaço vetorial é uma base para se escrever
qualquer estado possível do sistema. Desta forma, se caracterizarmos o
sistema por dois conjuntos de observáveis, não necessariamente compatíveis,
podemos expressar os estados resultantes de uma caracterização como uma
superposição de estados da outra e, desta maneira, poderá existir uma
transformação necessariamente unitária entre os espaços vetoriais associados
a cada caracterização, de modo a preservar a norma e outras características
geométricas do espaço. Conseqüentemente, as funções de transformação terão
grande importância na construção destas transformações unitárias, podendo,
assim, ser construída toda a caracterização cinemática da teoria.

Em um próximo trabalho demostraremos que podemos estabelecer a relação das
funções de transformação com a dinâmica de um sistema quântico, e como
pode-se estabelecer um princípio geral que é conhecido como Princípio
Variacional de Schwinger.

\noindent \textbf{\large Agradecimentos}

\bigskip

J. A. Ramirez agradece ao CNPq pelo suporte integral. B. M. Pimentel
agradece ao CNPq e à FAPESP pelo suporte parcial. C. A. M. de Melo agradece à
FAPEMIG pelo suporte parcial. Os autores estão em débito com o parecerista an%
ônimo pela revisão extremamente cuidadosa do artigo e por valiosas sugestões.

\end{document}